\documentclass[reprint, superscriptaddress, amsmath,amssymb, aps, prb, floatfix, longbibliography]{revtex4-2}

\usepackage{amsfonts}
\usepackage{braket}
\usepackage{dsfont}
\usepackage{graphicx}
\usepackage{xcolor}
\usepackage{listings}
\usepackage{comment}
\usepackage{hyperref}
\usepackage{booktabs} 

\usepackage{footnote}
\makesavenoteenv{figure}

\usepackage[caption=false]{subfig}

\begin{document}

\preprint{APS/123-QED}

\title{Machine-learned tuning to protected states by probing noise resilience}
    \author{Rodrigo A. Dourado}
\email{dourado.rodrigo.a@gmail.com}
\affiliation{ \textit{Departamento de Física, Universidade Federal de Minas Gerais, C. P. 702, 30123-970, Belo Horizonte, MG, Brazil}	}

\author{Nicol\'{a}s Mart\'{i}nez-Valero}
\affiliation{Instituto de Ciencia de Materiales de Madrid (ICMM), Consejo Superior de Investigaciones Cient\'{i}ficas (CSIC), Sor Juana In\'{e}s de la Cruz 3, 28049 Madrid, Spain}

\author{Jacob Benestad}
\affiliation{Department of Physics, Norwegian University of Science and Technology, Trondheim NO-7491, Norway}

\author{Martin Leijnse}
\affiliation{Division of Solid State Physics and NanoLund, Lund University, S-22100 Lund, Sweden}

\author{Jeroen Danon}
\affiliation{Department of Physics, Norwegian University of Science and Technology, Trondheim NO-7491, Norway}

\author{Rub\'{e}n Seoane Souto}
\affiliation{Instituto de Ciencia de Materiales de Madrid (ICMM), Consejo Superior de Investigaciones Cient\'{i}ficas (CSIC), Sor Juana In\'{e}s de la Cruz 3, 28049 Madrid, Spain}

\begin{abstract}
   Protected states are promising for quantum technologies due to their intrinsic resilience against noise. However, such states often emerge at discrete points or small regions in parameter space and are thus difficult to find in experiments. In this work, we present a machine-learning method for tuning to protected regimes, based on injecting noise into the system and searching directly for the most noise-resilient configuration. We illustrate this method by considering short quantum dot-based Kitaev chains which we subject to random parameter fluctuations. Using the covariance matrix adaptation evolutionary strategy we minimize the typical resulting ground state splitting, which makes the system converge to a protected configuration with well-separated Majorana bound states.
   We verify the robustness of our method by considering finite Zeeman fields, electron-electron repulsion, asymmetric couplings, and varying the length of the Kitaev chain. Our work provides a reliable method for tuning to protected states, including, but not limited to, isolated Majorana bound states.  
\end{abstract}
	
\maketitle

\textit{Introduction}. Encoding quantum information in so-called protected states, which are intrinsically immune to specific noise channels, represents one of the most promising avenues toward fault-tolerant quantum computation, since such protection can substantially enhance qubit coherence times and gate fidelities~\cite{Lidar2003,Gyenis2021,Danon2021}. 

In practice, protected qubit states can be realized in several ways.
The conceptually simplest approach is to tune the system to a sweet spot, either statically or dynamically, where the effective qubit Hamiltonian is, to leading order, insensitive to a dominant noise source.
This is a widely used technique to stabilize qubits and drastically improve their performance~\cite{Vion2002,Ithier2005,Koch2007,Martins2016,Reed2016,Sete2017,Nguyen2019,Didier2019,Frees2019,Huang2021,Didier2019a,Hong2020,Huang2021}.

More advanced protection strategies include employing topologically protected qubit encodings that are exponentially decoupled from all local perturbations~\cite{Nayak2008}.
A prominent example is given by the Majorana bound states (MBSs) that appear at the ends of one-dimensional topological superconductors~\cite{Kitaev2001}. These states could potentially be used as fundamental building blocks for topological quantum computation~\cite{kitaev2003fault, Nayak2008}, where their nonlocal nature and braiding-based operations provide inherent protection against local noise. The notion that such MBSs could be realized in spin–orbit-coupled, spin-polarized semiconductor nanowires proximitized by conventional $s$-wave superconductors sparked substantial efforts toward their experimental realization and detection~\cite{leijnse2012introduction, alicea2012new, DasSarma2015, aguado2017majorana, flensberg2021engineered, prada2020andreev, beenakker2020search, das2023search, kouwenhoven2024perspective, Souto_chapter}. An alternative approach consists of realizing a discrete lattice model of the one-dimensional topological superconductor (a so-called Kitaev chain~\cite{Kitaev2001}) in linear arrays of quantum dots (QDs) coupled via superconductors~\cite{Sau2012, Leijnse2012, liu2022tunable}, promising more direct control over the effective disorder potential.

For both approaches described above, navigating the many-dimensional parameter space to find the protected regime is often difficult and is typically performed using advanced tuning protocols~\cite{Aghaee_PRB2023} and/or machine-learning techniques~\cite{Tham_PRL2023, tham2024conductance, Benestad_PRB2024, van2024cross}. Especially when searching for topological protection, all of these approaches are hindered by the fact that the commonly used tuning metrics only \emph{indirectly} reflect the formation of a topologically non-trivial state. In the search for MBSs, one typically uses combinations of local and nonlocal conductance measurements~\cite{bolech2007observing, rosdahl2018andreev, Aghaee_PRB2023, tsintzis2022creating, Seoane2023, dourado2025measuringcoherencefactorsstates}, which, at best, provide limited evidence for their presence.

In this work, we propose a more direct approach to finding protected states, using their intrinsic noise resilience \emph{itself} as a tuning metric. We expect this to be a general strategy applicable to a broad class of platforms supporting different types of protected states. 
For concreteness, we focus on the example of QD-based Kitaev chains, motivated by the surge in experimental efforts in this direction~\cite{Dvir2023, ten2024two, haaf2024edgebulkstatesthreesite, bordin2024signatures, bordin2025probingmajoranalocalizationphasecontrolled, Bordin_arXiv2025, vanloo2025singleshotparityreadoutminimal, zhang2025gatereflectometryminimalkitaev}, as well as the complexity of the tuning problem~\cite{Benestad_PRB2024, van2024cross}.
Short Kitaev chains feature non-overlapping MBSs at discrete points in parameter space, known as MBS sweet spots. At these points, the two MBSs are localized at opposite ends, lie at zero energy, and are separated from higher excited states by a finite gap~\cite{Leijnse2012, tsintzis2022creating, luethi2024, Dourado_Kitaev3}. We show that one can tune to such MBS sweet spots by minimizing the ground-state energy splitting of a Kitaev chain subjected to local noise. We employ the covariance matrix adaptation evolutionary strategy (CMA-ES)~\cite{hansen2023cmaevolutionstrategytutorial}, using the resulting energy splitting of the Kitaev Hamiltonian due to random fluctuations of the QD levels as a loss function. We verify the generality of the method by varying the chain length, incorporating electron--electron interactions, and considering asymmetric setups.

\begin{figure}
\centering
\includegraphics[width=0.75\linewidth]{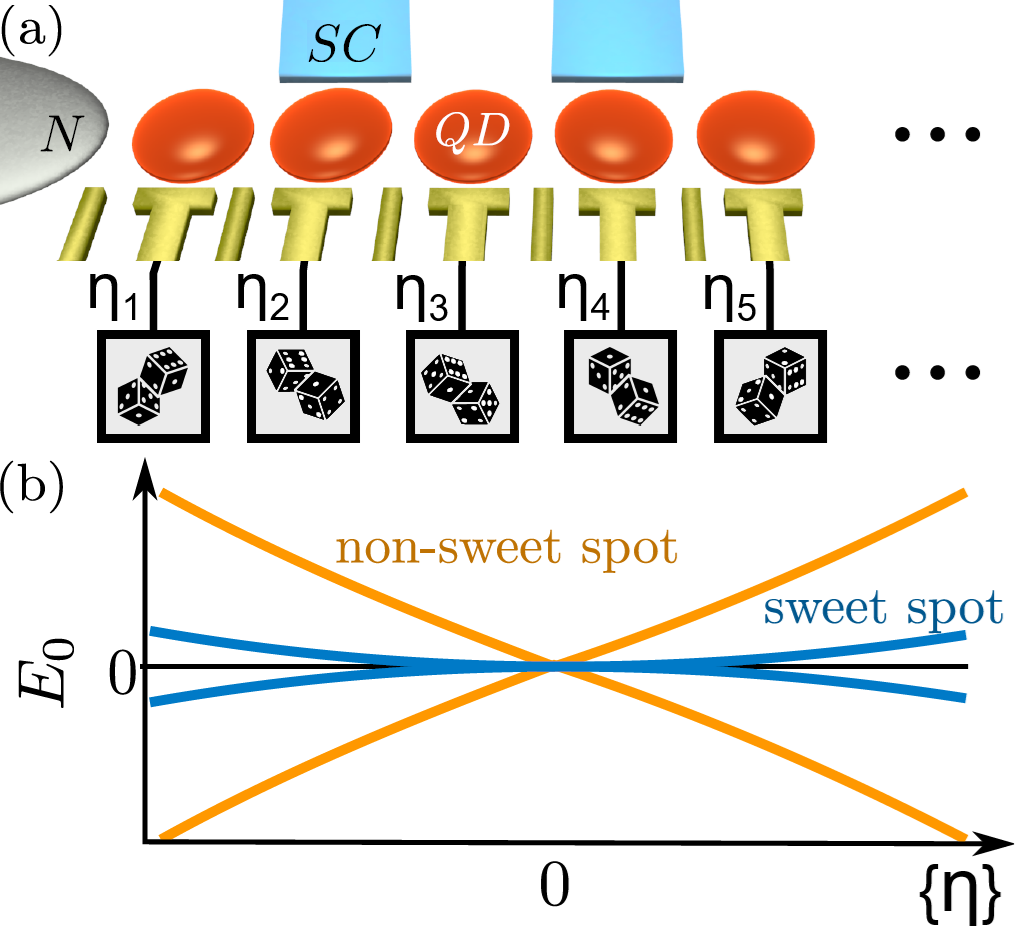}
\caption{Tuning to MBS sweet spots via parameter fluctuations. (a) Representation of a QD-based Kitaev chain with an arbitrary number of sites. We inject local noise ($\eta_i$) into the QDs as an optimization algorithm searches for the point in parameter space most stable against them. (b) Pictorial representation of the energy splitting $E_0$ as a function of the detunings represented by the set $\{\boldsymbol{\eta}\}$ for different points in parameter space. For a MBS sweet spot (in blue), $E_0$ remains robust (to a certain degree) upon variations in the QD levels, while for non-sweet spots (in orange), the degeneracy of the ground state splits.}
\label{Fig1}
\end{figure}

\textit{Quantum dot-based Kitaev chains}. We consider an $n_s$-site Kitaev chain hosted in an array of $N=2n_s-1$ QDs, see Fig.~\ref{Fig1}(a). All even QDs are strongly proximitized by superconductors (blue), yielding Andreev bound states, whose energy is controlled by the superconducting QD levels, and that mediate elastic cotunneling (ECT) and crossed Andreev reflection (CAR) processes between the normal (odd) QDs.
The electronic states on the QDs are described by the following model Hamiltonian~\cite{liu2022tunable,tsintzis2022creating}
\begin{equation}
\begin{split}
    H_{0} &= \sum_{\substack{i=1\\ \sigma = \uparrow,\downarrow}}^N \left(\varepsilon_{i} + s_\sigma \frac{V_{z, i}}{2} \right) n_{i, \sigma}
    +\sum_{i=1}^N \left(\Delta_i c_{i, \uparrow}^\dagger c_{i, \downarrow}^\dagger + {\rm H.c.} \right),
    \label{Eq:Hmicroscopic}
\end{split}
\end{equation}
where the operator $c_{i,\sigma}$ annihilates an electron with spin $\sigma$ on dot $i$ and $n_{i,\sigma} = c^\dagger_{i,\sigma}c_{i,\sigma}$.
The $\varepsilon_{i}$ are the QD onsite potentials and $V_{z,i}$ the Zeeman energies, with $s_\sigma=\pm1$ for spin $\sigma = {\uparrow,\downarrow}$. The superconductivity is described by an induced $s$-wave pairing amplitude in the even QDs, {\it i.e.}, $\Delta_i = 0 \, (\Delta)$ for $i$ odd (even). The tunneling between neighboring QDs is described by
\begin{equation} \label{TunnelingQDs}
    H_{\rm T}= \sum_{\substack{i=1\\\sigma = \uparrow,\downarrow}}^{N-1}\left( t_i c_{i+1, \sigma}^\dagger c_{i, \sigma} + t_{i}^{so} s_{\sigma}c_{i+1, \sigma}^\dagger c_{i, \bar{\sigma}}+  {\rm H.c.} \right),
\end{equation}
where the $t_i$ are the spin-conserving hopping amplitudes and $t^{so}_i$ the spin-flip tunneling amplitudes due to spin--orbit coupling, where $\bar{\sigma}$ is the opposite spin to $\sigma$. The total Hamiltonian is given by $H = H_{0} + H_{\rm T}$. In the following, we set $t_i = 0.5\Delta$, $t_{i}^{so} = 0.2\Delta$, 
and $V_{z,i} = 1.5\Delta \, (V_{z,i} = 0)$ for $i$ odd (even), accounting for screening effects due to the strong coupling to superconductors. We emphasize that our qualitative results do not depend on this specific choice of parameters. We neglect the effects of electron–electron interactions, which allows us to work in the Bogoliubov–de Gennes (BdG) representation. We have verified that including interactions does not qualitatively change the results [see Sec.~C of the Supplementary Material (SM)].

\textit{Tuning Kitaev chains to protected states}. MBS sweet spots are characterized by three properties: (i) a ground state degeneracy, i.e., $E_0 = 0$, where $E_0$ is the energy difference between the lowest-energy states in the even and odd parity sectors; (ii) a finite gap to the other states, $E_{\rm ex} = E_1 - E_0$, where $E_1$ is the first energy above $E_0$; (iii)  highly localized MBSs on opposite ends of the chain. We quantify the MBS overlap at the outermost QDs, $i=1,N$, using the Majorana polarization (MP)~\cite{sticlet2012spin, sedlmayr2015visualizing, sedlmayr2016majorana, tsintzis2022creating, Awoga_PRB2024}
\begin{equation} \label{MPdefinition}
    M_i = \frac{\sum_{ \sigma}2 u_{i, \sigma}v_{i, \sigma}}{\sum_{ \sigma} u_{i, \sigma}^2 + v_{i, \sigma}^2},
\end{equation}
where $u_{i, \sigma}$ and $v_{i, \sigma}$ are the local coherence factors for the lowest energy mode, obtained from the electron- and hole-components of the BdG wave function in the non-interacting limit.

A characteristic and defining feature of these sweet spots is the robustness of $E_0$ against parameter fluctuations. For instance, a Kitaev chain with $n_s$ sites tuned to a sweet spot will preserve the ground state degeneracy upon variations of the energies of $n_s-1$ sites. This is pictorially shown in Fig.~\ref{Fig1}(b). For practical realizations of Kitaev chains at finite Zeeman splittings, as in Eq.~(\ref{Eq:Hmicroscopic}), this robustness is only approximate for all sweet spots~\cite{Dourado_Kitaev3}. These extra layers of imperfections motivate our search for a tuning protocol that leads to the highest degree of protection in QD-based Kitaev chains.

In the following, we exploit this robustness to develop a method to tune Kitaev chains to sweet spots featuring well-localized MBSs, by applying an automated optimization to minimize the sensitivity of $E_0$ to fluctuations of the onsite potentials $\varepsilon_i$.
While it is possible to tune an $n_s$-site Kitaev chain by probing its resilience to fluctuations only of the $n_s-1$ normal-site potentials, we go one step further and apply random fluctuations to all $N=2n_s-1$ QD levels. By randomly varying all QD levels, including those on the superconducting QDs, we search for the points that are most robust against fluctuations in all effective parameters of the Kitaev chain: the onsite levels as well as the ECT and CAR amplitudes. 

We employ the CMA-ES optimization algorithm~\cite{hansen2023cmaevolutionstrategytutorial} in a numerical simulation of such a tuning procedure.
For each step of the simulation, the algorithm draws from a multivariate normal (MvN) distribution a population of $n_{\rm pop}=20$ tuning configurations (i.e., sets of onsite potentials) $\boldsymbol\varepsilon^{(i)}=\left\{\varepsilon^{(i)}_1, \varepsilon^{(i)}_2, \dots, \varepsilon_N^{(i)} \right\}$, where $1\leq i \leq n_{\rm pop}$. Each configuration is then assessed via a loss function based on a numerical evaluation of $E_0$,
\begin{equation}
    L \left(\boldsymbol\varepsilon^{(i)} \right) = \frac{1}{P} \sum_{p=1}^P E_0 \left(\boldsymbol\varepsilon^{(i)} + \boldsymbol\eta^{(p)} \right),\label{eq:loss}
\end{equation}
where the $P=200$ vectors $\boldsymbol\eta^{(p)} = \left\{\eta^{(i)}_1, \eta^{(i)}_2, \dots, \eta^{(i)}_N \right\}$ represent the fluctuations, each element randomly drawn from uniform distribution in the interval $[-W,W]$, where we set $W = 0.075\Delta$~\footnote{This value is roughly half of the excitation gap for 2-site Kitaev chains~\cite{tsintzis2022creating}. The amplitude $W$ controls the optimization of $E_0$ and the MP, with larger values penalizing MBS overlaps more heavily, thereby increasing the MP, but also increasing the final value of $E_0$.}.
The $n_{pop}/2$ configurations that yield the lowest loss are then used to update the parameters of the (MvN) distribution, after which the next generation of $n_{\rm pop}$ tuning configurations is drawn. 
The optimization procedure terminates when the loss of all members of the population becomes smaller than a defined threshold, see Sec.~A of the SM for more details of the simulations. 

In this work, we focus on the noise resilience of zero-energy states by including a sensitivity to the magnitude of $E_0$ in the loss function, Eq.~(\ref{eq:loss}). However, our method is general and can be extended to states with non-zero energy, for instance by defining the loss function as $L(\boldsymbol\varepsilon^{(i)}) = \{\sum_{p=1}^P [ E_0(\boldsymbol\varepsilon^{(i)} + \boldsymbol\eta^{(p)}) - E_0(\boldsymbol\varepsilon^{(i)}) ]^2/P \}^{1/2}$, which exclusively aims for maximization of the resilience of $E_0$ to noise.

To verify the robustness of our method, we perform $50$ independent optimization runs. During each run, we monitor the best tuning configuration at each generation, i.e., the one that minimizes the loss. For these ``optimal'' configurations, we calculate the energy splitting $E_0$, the MP at the outer QDs $M_{1,N}$, and the excitation gap $E_{\rm ex}$. For each quantity $x$, at each generation, we take the median value (across the $50$ simulations) and calculate the standard deviations above and below the median, $\sigma_x^{\pm} = \{\sum_{x_m \gtrless \Bar{x}} (\Bar{x} - x_m)^2/25\}^{1/2}$, where $\Bar{x}$ represents the median value.
In the following, we present the medians and standard deviations over the generations for all tuning procedures.

\begin{figure}
\includegraphics[width=\linewidth]{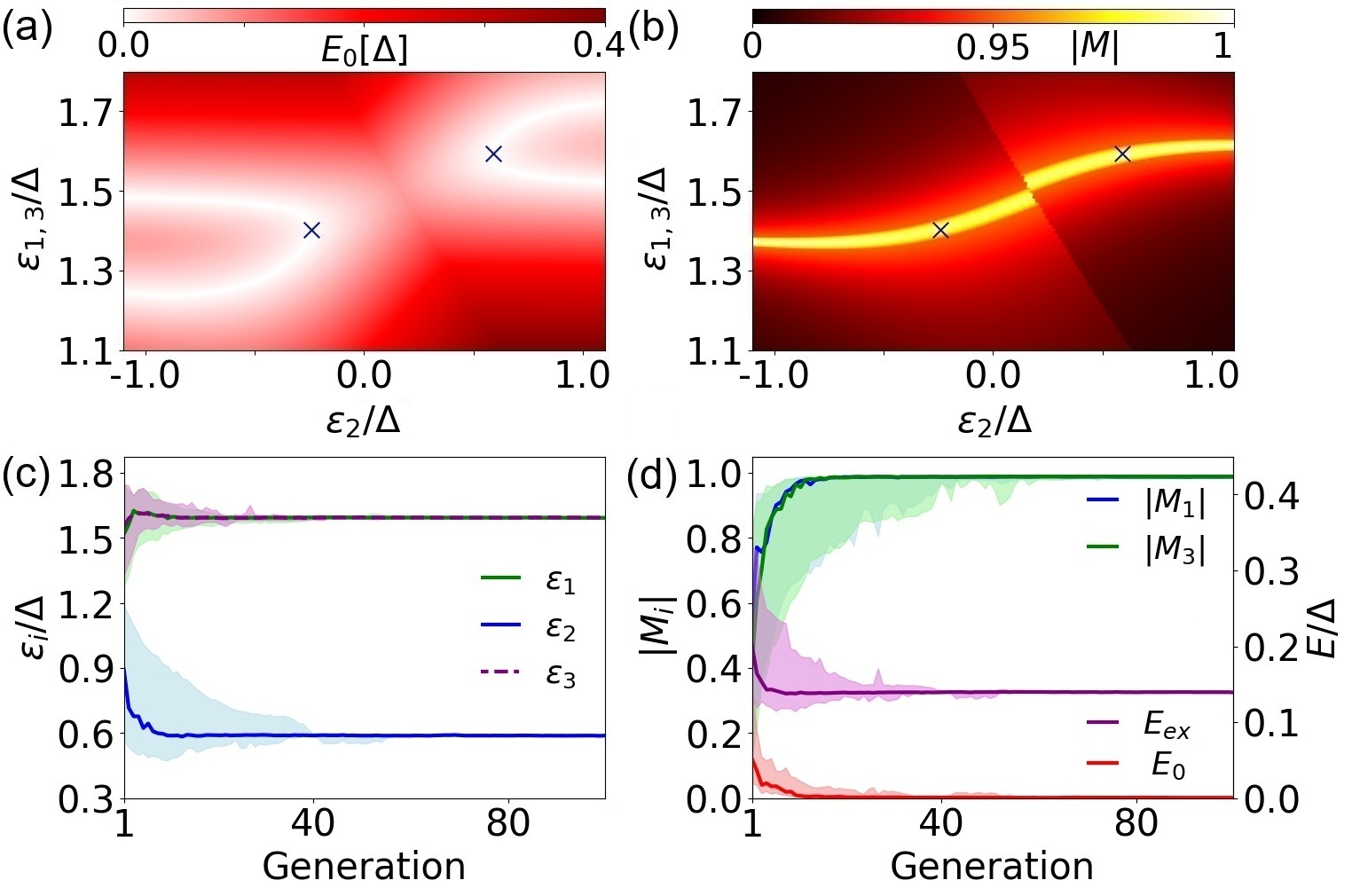}
\caption{Results for the 2-site Kitaev chain. (a,b) Energy splitting $E_0$ and MP at the outer QDs, as a function of the normal and superconducting QD levels, $\varepsilon_{1, 3}$ and $\varepsilon_{2}$, respectively. The blue crosses indicate sweet spots found by the automated tuning procedure. (c,d) Median values (lines) and standard deviations (shades) of (c) the three QD potentials and (d) MBS properties, i.e., MPs $M_1$ (blue) and $M_3$ (green), $E_{\rm ex}$ (purple), and $E_0$ (red), corresponding to the best members of each generation across $50$ independent simulations.}
\label{Fig2}
\end{figure}

\textit{2-site Kitaev chains}. We first demonstrate that the tuning protocol leads to MBS sweet spots in a 2-site Kitaev chain ($N=3$), the shortest system that can support localized MBSs. 
The results are shown in Fig.~\ref{Fig2}. As a reference, we show in Figs.~\ref{Fig2}(a,b) the calculated energy splitting $E_0$ between the even- and odd-parity ground states and the MP on the outer QDs, as a function of $\varepsilon_1 = \varepsilon_3 = \varepsilon_{1, 3}$ ($|M_{1}| = |M_3| = |M|$) and $\varepsilon_2$. The MBS sweet spots can be found at intersections between the white lines in Fig.~\ref{Fig2}(a), which represent degeneracy of the ground state, and the high MP lines in Fig.~\ref{Fig2}(b), where the MBSs are localized at the outer QDs.
In Figs.~\ref{Fig2}(c,d), we show the statistics of 50 independent simulations where the CMA-ES algorithm tunes the vector $\boldsymbol\varepsilon$ using the loss function defined in Eq.~(\ref{eq:loss}). For each of the $50$ independent simulations, the initial MvN distribution for the onsite potentials has a standard deviation of $\Delta$ and the mean for each variable is randomly drawn from the intervals $\Delta \leq \varepsilon_{1},\varepsilon_{3} \leq 2\Delta$ (here, $\varepsilon_1$ and $\varepsilon_3$ are treated as independent variables) and $0.2\Delta \leq \varepsilon_2  \leq 1.2\Delta$, which also bound the search area. This randomness reflects
natural variations in experiments.
The medians of the QD levels in Fig.~\ref{Fig2}(c) and of the energy splitting $E_0$, the excitation gap $E_{\rm ex}$, and the MPs $|M_{1}|$ and $|M_3|$ in Fig.~\ref{Fig2}(d) are plotted as solid lines, while the respective standard deviations are represented by the shades. 
We note that in all 50 cases the simulations converged to the same point in parameter space, marked by the right blue cross in Figs.~\ref{Fig2}(a,b).
When we perform the same simulations, but choosing $-1.2\Delta \leq \varepsilon_{2}\leq 0.2 \Delta$, we consistently end up at the point marked by the left cross.
Using less restrictive bounds results in randomly varying convergence to either of the points.

Figure~\ref{Fig2}(d) shows that the tuning indeed always converges to a point where the energy splitting $E_0$ is very small (red), the two MBSs become isolated at the ends, indicated by the quantities $|M_{1}|, |M_3|\approx 1$ (blue and green, respectively), and the excitation gap is sizable (purple). 
This is indeed a MBS sweet spot of the 2-site chain and, since we found it by maximizing resilience to local noise, this confirms that the sweet spot is also the point where $E_0$ is least sensitive to fluctuations.
In long Kitaev chains, where topological phases are well-defined, the degeneracy of the ground state is robust against local perturbations~\cite{Kitaev2001}, and we see that this notion extends to short Kitaev chains, where MBS sweet spots are also associated with the highest degree of protection.

A more quantitative overview of the results presented in Figs.~\ref{Fig2}(c,d) is presented in the tables in Sec.~B of the SM.
We have verified that the presented results also hold when including intra-QD interactions and for asymmetric QD arrays, where the hoppings and spin--orbit couplings between the superconducting and normal QDs are different, which is usually the case in experiments~\cite{Dvir2023, bordin2024signatures, ten2024two, bordin2025probingmajoranalocalizationphasecontrolled}, see Sec.~C of the SM.

\begin{figure}
\centering
\includegraphics[width=\linewidth]{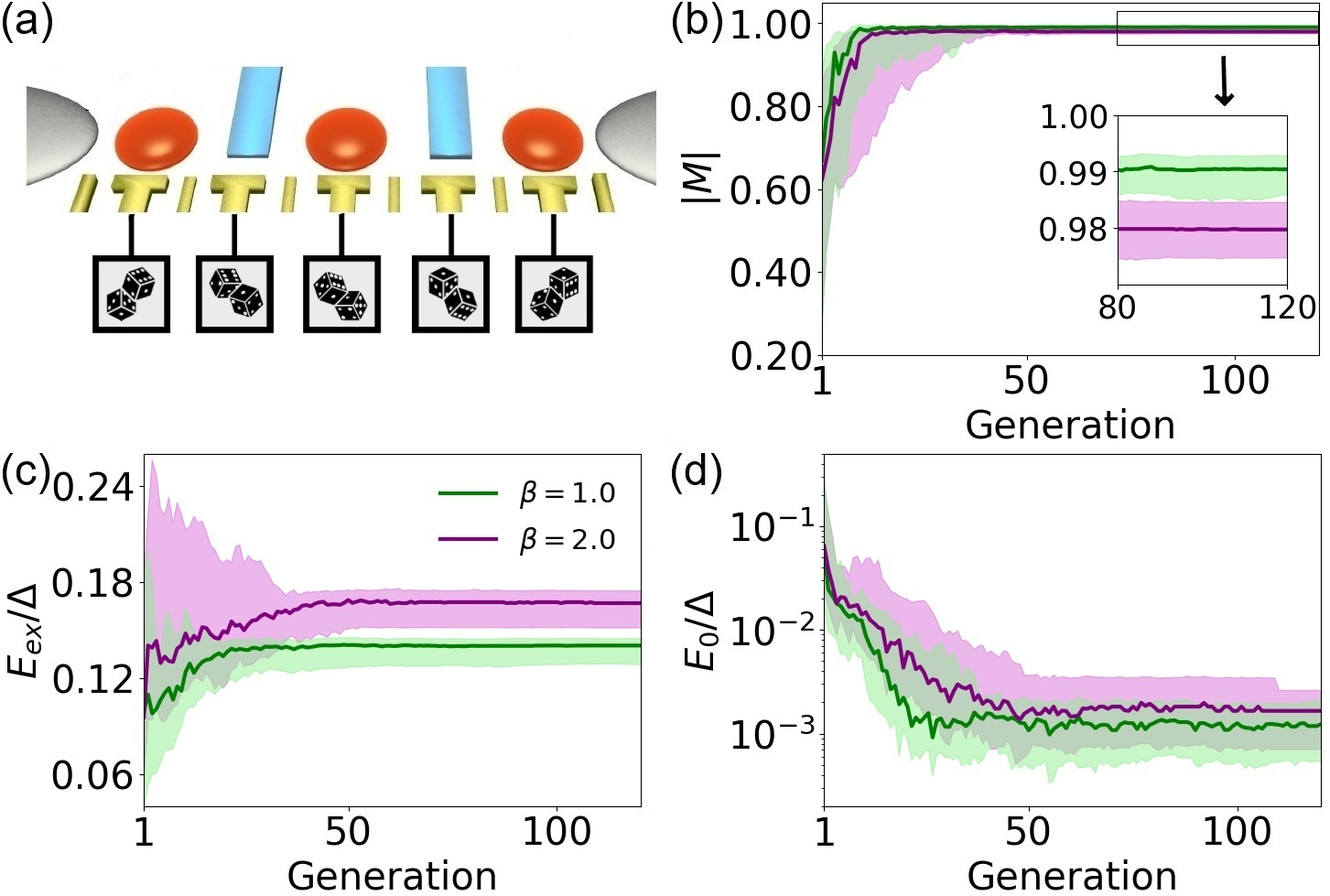}
\caption{Results for the 3-site Kitaev chain. (a) Sketch of the system. 
(b--d) Median values (lines) and standard deviations (shades) of the MBS properties [(b) MP $|M_1| = |M_5| = |M|$, (c) $E_{\rm ex}$, and (d) $E_0$] corresponding to the best members of each generation across $50$ independent simulations.
The simulations were performed with $\beta = 1$ (green) and $\beta=2$ (purple). All parameters are the same as in Fig.~\ref{Fig2}, apart from $W = 0.1\Delta$.}
\label{Fig3}
\end{figure}

\textit{3-site Kitaev chains}. We next analyze the 3-site Kitaev chain, where an additional pair of superconducting and normal QDs is attached to the 2-site chain ($N = 5$), as shown in Fig.~\ref{Fig3}(a).
Increasing the chain length can improve the ground state protection at the sweet spots~\cite{bordin2024signatures, Dourado_Kitaev3}, and gives rise to a wide variety of MBS sweet spots that differ in the MBS localization and excitation gap. In particular, the strong renormalization experienced by the central QD (that couples two superconducting QDs) leads to an effective energy shift with respect to the outermost ones (that only couple to one superconducting QD). This shift can increase the MP but reduces the excitation gap, since detuning the central QD from resonance suppresses the ECT and CAR amplitudes, and causes the ground-state to split quadratically when the outermost QDs are detuned simultaneously~\cite{Dourado_Kitaev3}. The latter feature allows us to control the characteristics of the sweet spots by adjusting the noise in the outermost QDs, selecting $\eta_{1,5}$ from the interval $[-\beta W, \beta W]$, while the other fluctuation parameters are drawn from $[-W, W]$, as before. In this way, $\beta > 1$ more strongly penalizes sweet spots where the central QD is detuned, thereby reducing the formation of barriers within the QD array and increasing the excitation gap.

The above discussion is corroborated by our simulations for the 3-site chain with different values of $\beta$, as presented in Fig.~\ref{Fig3}.
For $\beta = 1$, the simulations converge to a sweet spot with $|M| \approx 0.99$ and $E_{\rm ex} \approx 0.14 \Delta$, see the green lines in Figs.~\ref{Fig3}(b,c). By setting $\beta = 2$, the algorithm converges to sweet spots with larger excitation gaps, $E_{\rm ex} \approx 0.17 \Delta$, but slightly reduced MP, $|M| \approx 0.98$, see the purple lines in Figs.~\ref{Fig3}(b,c). This illustrates the trade-off between MP and the excitation gap.

\begin{figure}
\centering
\includegraphics[width=\linewidth]{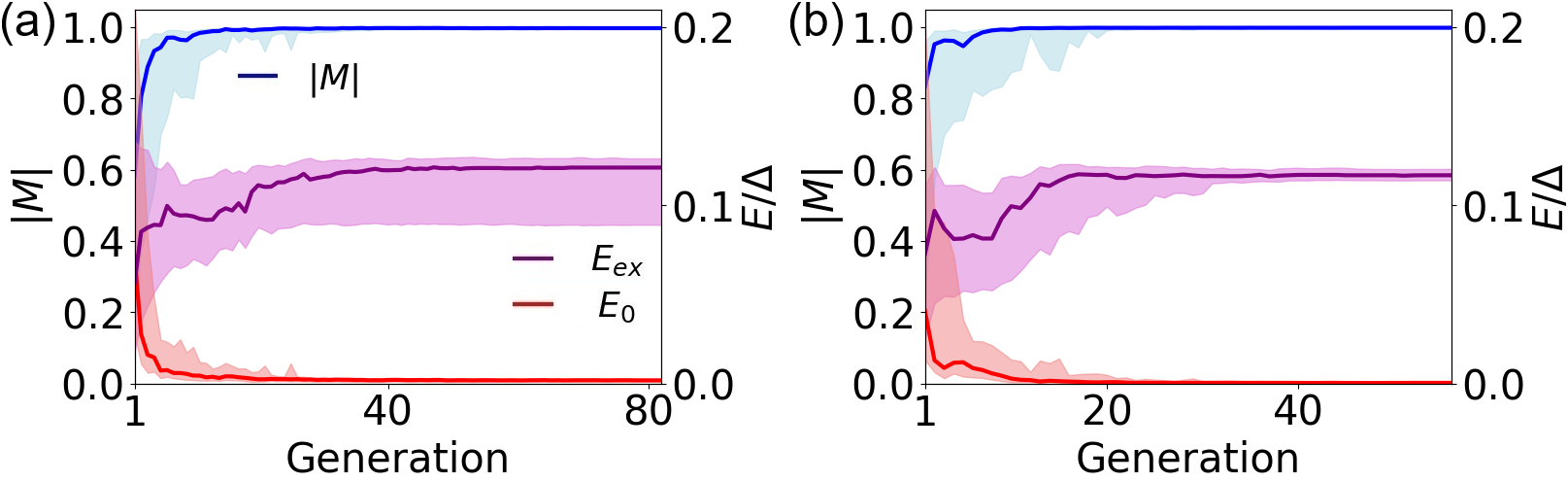}
\caption{Results for longer Kitaev chains. 
Evolution of the median values (lines) and standard deviations (shades) of $|M_{1}| = |M_N| = |M|$ (blue), $E_0$ (red), and $E_{\rm ex}$ (purple) corresponding to the best members of each generation across $50$ independent simulations, for (a) 4- and (b) 5-site Kitaev chains.
All parameters are the same as in Fig.~\ref{Fig3}, apart from $W= 0.125\Delta$.}
\label{Fig4}
\end{figure}

\textit{Longer Kitaev chains}. We also verify the applicability of our tuning protocol to longer chains. As the number of QDs increases, it becomes more challenging to fine-tune phases between the superconductors and ensure the alignment of all chemical potentials, which can lead to reductions of the excitation gap~\cite{Sau2012, Dourado_Kitaev3, chunxiao2025scaling, bordin2025probingmajoranalocalizationphasecontrolled}. However, we verify that the correspondence between minimizing the energy splittings due to fluctuations and MBS sweet spots is maintained in longer chains. We illustrate this for Kitaev chains with $4$ ($N = 7$) and $5$ ($N = 9$) sites in Figs.~\ref{Fig4}(a,b), respectively. In longer chains, increasing the amplitude of the fluctuations at the outermost QDs through the parameter $\beta$ is no longer a reliable method of enhancing the excitation gap. The excitation gap also does not always converge to the same values, see the purple shades in Figs.~\ref{Fig4}(a,b), in contrast to smaller chains, Fig.~\ref{Fig2}(d) and Fig.~\ref{Fig3}(c). A possible way of overcoming these limitations could be to include direct measurements of the excitation gap into the loss function. 

\textit{Conclusions}. In this work, we propose a general method of tuning to protected states by directly probing their resilience against local noise.
As an example, we demonstrate how to tune to MBS sweet spots by minimizing the average energy splitting between the even and odd ground states in QD-based Kitaev chains when subjected to random parameter fluctuations. We test our proposal in 2-site Kitaev chains and show that minimizing the energy splitting due to random fluctuations on the QD levels leads to sweet spots with well-localized MBSs. We also verify the robustness of the method by considering electron--electron interactions, longer chains, and asymmetric setups. Our method can be readily applied to experiments, since the only required measurement is the ground-state energy splitting, $E_0$.
This could be done, e.g., via conductance spectroscopy measurements, which are limited by thermal broadening, or in a qubit setup using Ramsey-type experiments~\cite{Tsintzis2024, pan2025rabi}.
We emphasize that the method is not limited to the case of artificial Kitaev chains, but should be applicable to find protected states of any topological or non-topological nature.

\textit{Acknowledgments}. R.A.D. acknowledges financial support from Coordenação de Aperfeiçoamento de Pessoal de Nível Superior (CAPES), Brazil (Grant No. 88887.111639/2025-00). M.L. acknowledges funding from the European Research Council (ERC) under the European Unions Horizon 2020 research and innovation programme under Grant Agreement No. 856526, the Swedish Research Council under Grant Agreement No. 2024-05491, and NanoLund. R.S.S acknowledges funding from the Horizon Europe Framework Program of the European Commission through the European Innovation Council Pathfinder Grant No. 101115315 (QuKiT), the Spanish Comunidad de Madrid (CM) ``Talento Program'' (Project No. 2022-T1/IND-24070), and the Spanish Ministry of Science, innovation, and Universities through Grants PID2022-140552NA-I00.

\bibliography{bibliography.bib}

\end{document}